\documentclass[11pt,a4paper]{article}
\usepackage{jcappub}
\usepackage{graphicx}
\newcommand{\FLRW}{{\it FLRW\/}}
\title{Dynamics of Void and its Shape in Redshift Space}

\author[a]{Kei-ichi  Maeda}
\author[b]{,Nobuyuki Sakai}
\author[c]{,Roland Triay}

\affiliation[a]{Department of Physics \& RISE,  Waseda University,\\
Okubo 3-4-1, Shinjuku, Tokyo 169-8555, Japan}
\affiliation[b]{Department of Education, Yamagata University,\\
Yamagata 990-8560, Japan}
\affiliation[ c]{Centre de Physique Th\'eorique\footnote{Unit\'e Mixte de Recherche (UMR 6207) du CNRS, et des universit\'es  Aix-Marseille I, Aix-Marseille II et du Sud Toulon-Var. Laboratoire affili\'e \`a la FRUMAM (FR 2291).},\\
CNRS Luminy Case 907, 13288 Marseille Cedex 9, France}

\emailAdd{maeda@waseda.jp}
\emailAdd{nsakai@e.yamagata-u.ac.jp}
\emailAdd{triay@cpt.univ-mrs.fr}

\abstract{
We investigate the dynamics of a single spherical void embedded in a  Friedmann-Lema\^itre universe, and analyze the void shape  in the redshift space. We find that the void in the redshift space appears as  an ellipse shape elongated in the direction of the line of sight ({\em i.e.\/}, an opposite deformation to the Kaiser effect). Applying this result to observed void candidates at the redshift $z\sim 1$-2, it may provide us with a new method  to evaluate the cosmological parameters, in particular the value of a cosmological constant.
}


\arxivnumber{...}

\begin{document}
\maketitle 

\section{Introduction}
From the observational data (Type Ia supernovae \cite{dark_energy1}, the cosmic microwave background (CMB)  \cite{dark_energy2}, galaxy distribution \cite{dark_energy3}, and weak lensing \cite{dark_energy4}), we find that  the $\Lambda$CDM model is the most generally accepted approach in cosmology. It is based on the Friedmann-Lema\^itre solution of the Einstein  equations with a cosmological constant $\Lambda$, which describes the universe with a uniformly distributed dust.  The observation also reveals the existence of large scale structures, i.e., the large voids with the scale of about $100$Mpc as well as the 
clusters of galaxies.  It is important to explain how such cosmological large scale structures are formed from the initially very small density fluctuations at the recombination epoch. The time evolution of fluctuations depends on the background universe, which is described by cosmological parameters such as the Hubble constant $H_0$, the density parameter $\Omega_0$, and a cosmological constant $\Lambda$. Hence, the observations of the large scale structures as well as 
the CMB anisotropies give the constraints on the parameters. There are many works on the formation of large scale structures and their observational properties related to cosmological parameters. However, they are mostly concerned with over-densities, such as clusters of galaxies. Their dynamical characteristics act on their appearance in the redshift space. For example, because of the Doppler shift associated with the peculiar velocities of galaxies, the velocity dispersion within a cluster makes its shape elongated toward the observer, and hence it is named ``Finger of God"\cite{finger-of-god-effect}. The Kaiser effect is also caused by peculiar velocities and arises from coherent motions as the galaxies fall inwards towards the cluster center \cite{Kaiser_effect}. The Kaiser effect usually leads an apparent flattening of the structure, in accordance with the particular dynamics of the situation. 

On the other hand, the under-dense regions such as voids have not been intensively investigated as large scale structures as it has been done for over-dense regions. A few models of voids have been investigated (see {\em e.g.\/} \cite{origins}, and references therein) in order to interpret observations at high redshift. The hot and cold spots detected in the CMB  may be associated with super-clusters and huge voids, which are identified in the Sloan Digital Sky Survey catalog \cite{GNS}. Subsequently, the integrated Sachs-Wolfe effect\cite{SWeffect} due to spherically symmetric super-structures suggested the presence of nonlinear voids or clusters with a  radius $100\sim200h^{-1}$Mpc \cite{IST}.\footnote{One interesting proposal for ``dark energy" is  an inhomogeneous universe model without any  exotic ansatz such as introduction of dark energy or modification of gravity \cite{Tom,Alnes}. In this model, we are living near the center of  the under-dense region (a huge void). This model is fascinating if it turns out to be able to explain all observational constraints including the CMB anisotropy,although it is against the Copernican world view.}

Herein, we focus on the under-dense regions  that undergo cosmological expansion with the aim to investigate the effect of the cosmological parameters on their dynamics. As a first step of our investigation for such under-dense regions, we discuss the case of a single isolated void embedded in the uniform and isotropic 
universe, which is the most simple example to understand.

In \S.\ref{basic_eq}, we briefly introduce our model with the basic equations for a single void dynamics based on general relativity (GR). Our analysis on the  expansion of the void is given in \S.\ref{void_dynamics}, by which we confirm the previous result based on a covariant formulation of Newtonian dynamics \cite{Fliche_Triay}.  In \S.\ref{redshift_space}, we analyze the shape of a void in the redshift space by which we show that such a distortion effect could be observable, similarly to  ``Fingers of God'' or to the Kaiser effect. Conclusion follows in \S.\ref{Conclusion}.

\section{Modeling a spherical void in the expanding universe}
\label{basic_eq}
In this paper, we consider a single isolated void in GR to investigate its dynamics. We propose a model of a spherical void in the universe by using the thin-shell approximation based on the Israel's formalism \cite{Isr}. A void in our model is described by an under-dense region filled by a uniform distribution of dust, which is embedded in the Friedmann-Lema\"{\i}tre (FL) universe with a positive cosmological constant $\Lambda\geq 0$. Here we will analyze only the case of an empty void. 
The extension to the case of a non-empty void is straightforward (see Appendix \ref{appendix_A}). The equations of motion for a void were originally derived by Maeda and Sato \cite{MS}, and later they were rewritten in a more convenient expression \cite{SMS} that we will use here.

\subsection{Basic Equations}
We assume that a void is characterized by a spherical shell $S$ of the radius $R$ embedded in an expanding uniform background universe. It defines the boundary layer between two spatial regions filled by two different isotropic and uniform distributions of dust. For convenience in writing, we use the indices ``$-$'' and ``$+$'' for the variables related to the inner region and to the outer region, respectively. When no ambiguity is present,  we will use the symbol ``$\pm$'' in order to write two equations as a single one. We assume that the corresponding space-times are described by a Friedmann-Lema\^itre-Robertson-Walker (FLRW) metric
\begin{eqnarray}\label{FLRW}
{\rm d}s_\pm^{2}=-{\rm d}t_\pm^{2}+a_\pm^{2} 
\left[{\rm d}\chi_\pm^{2}+f_\pm^{2}(\chi_\pm) {\rm d}\Omega^{2}
\right],\qquad 
f_\pm(\chi_\pm)= \left \{
\begin{array}{ll}
 \sin{\chi_\pm} &(k_\pm=1 ) \\
 \chi_\pm &(k_\pm=0 ) \\
 \sinh{\chi_\pm} &(k_\pm=-1)
\end{array}
\right.
\end{eqnarray}
where $k_\pm$ are the signs of the scalar curvatures and $a_\pm$ the scale factors  of the FLRW space-times. Both space-times satisfy the Friedmann equations
\begin{eqnarray}\label{Friedmann}
H_\pm^{2}+\frac{k_\pm}{ a_\pm^{2}}=\frac{8\pi G}{ 3} \rho_\pm,\qquad
{\rm with}\qquad
H_\pm=\frac{\dot{a}_\pm}{a_\pm},
~~{\rm and}~~
 \rho_\pm=\rho_{\rm vac}+\rho_\pm^{\rm (m)}
\end{eqnarray}
where the dotted variables denote their time derivatives, $\rho_\pm^{\rm (m)}\propto a_\pm^{-3}$ are the matter densities, and $\rho_{\rm vac}= \Lambda/(8\pi G)$ accounts for a cosmological constant (viewed as a contribution to gravity in term of energy density). These space-time regions are joined through $\Sigma$, the three-dimensional{\em world volume\/} of $S$ in the space-time ({\it i.e.}, the hyper-surface that corresponds to the envelope of $S$ evolving over time), which metric reads
\begin{eqnarray}\label{3metric}
{\rm d}s_{\Sigma}^{2}=-{\rm d}\tau^{2}+R^{2}(\tau) {\rm d}\Omega^{2}
\end{eqnarray}
where $\tau$ is the proper time of $S$. The continuity of the space-time requires the following conditions to hold
\begin{eqnarray}
&&R=a_{+} f_{+}\left(\chi_{+}^{(\Sigma)}\right)=a_{-} f_{-}
\left(\chi_{-}^{(\Sigma)}\right)\\
&&d\tau^2=dt_+^2-a_+^2(t_+)\left(d\chi_+^{(\Sigma)}\right)^2 =
dt_-^2-a_-^2(t_-) \left(d\chi_-^{(\Sigma)}\right)^2
\end{eqnarray}
These imply the following equations; 
\begin{eqnarray}
&& \frac{dR }{ dt_+} =f_{\pm}'\left(\chi_{\pm}^{(\Sigma)}\right) v_{\pm}
 +H_{\pm} R,\quad
v_{\pm} \equiv a_{\pm} \frac{d\chi_{\pm}^{(\Sigma)} }{ dt_{\pm}},\quad
f^{\prime}(\chi_{\pm})= \sqrt{1-\frac{k_{\pm}R^{2}}{a_{\pm}^{2}}}
\label{eom1}\\
&&\gamma_+(f'_+v_+ + H_+R) =\gamma_- (f'_-v_- + H_-R)
\label{eom2}\\
&& \frac{dt_-}{ dt_+}= \frac{\gamma_-}{\gamma_+},\quad
\gamma_{\pm}\equiv\frac{1}{\sqrt{1-v_{\pm}^{2}}}
\,. 
\label{eom3}
\end{eqnarray}
where $v$ is the peculiar velocity\footnote{It stands for the the radial component of the peculiar velocity field, but we use such a shorter terminology thanks to its spherical symmetry.} of $S$ and $\gamma$ the Lorentz factor. The jump condition on the extrinsic curvature of $\Sigma$ provides us with the following basic equations;
\begin{eqnarray}\label{eom4}
&& \frac{d(\gamma_+v_+)}{ dt_+}=
-\gamma_+v_+H_++ \frac{Gm_S}{^2}-\frac{[4\pi R^2\gamma^2v^2\rho^{\rm (m)}]^{\pm}
}{ m_S}\\
&&\gamma_+ \frac{d m_S}{ dt_+}=[4\pi R^2\gamma^2v\rho^{\rm (m)}]^{\pm}
\label{eom5}\\
&&\gamma_+[\gamma(f'+v HR)]^{\pm}= -\frac{Gm_S}{ R}
\,,
\label{eom6}
\end{eqnarray}
where $m_S=m_S(t_+)$ is the mass of the shell and the bracket $[~]^{\pm}$ stands for $[\Psi]^{\pm}\equiv\Psi_+-\Psi_-$ (see \cite{Isr,MS,SMS} for details).

The dynamics of the shell $S$ is defined by Eqs.\,(\ref{eom1})-(\ref{eom5}) with the constraint (\ref{eom6}). In the case of an empty void ($\rho_-^{\rm (m)}=0$), we do not have to solve Eqs.\,(\ref{eom2}) and (\ref{eom3}), which give the information for $v_-$ and $t_-$. We use the constraint (\ref{eom6}) to give an initial value of $m_S$ as well as to check the precision of the numerical integration schema.

\subsection{Model parameters}\label{Mparameters}
Since the outer region characterizes the background universe, the index ``$+$''will be omitted, which enables us to retrieve the notations used in cosmology ({\em e.g.\/}, the Hubble parameter of the universe $H$ stands for the expansion rate of the outer region, and so on.

By setting the present value of the scale factor $a_{0}=1$, the redshift of an event which has happened at time $t$ and  is observed today ($t=t_{0}$) is given by $z=1/a(t)-1$. The integration of the above differential equations requires initial values for the variables
\begin{equation}\label{parameters}
R,\quad v,\quad {\cal H}^-= \frac{H^-}{ H},\quad
\delta^-\equiv \frac{\rho_-^{\rm (m)}}{\rho^{\rm (m)}}-1,\quad 
\Omega\equiv \frac{8\pi G\rho^{\rm (m)}}{ 3H^2},\quad {\rm and } \quad
\lambda\equiv \frac{\Lambda}{ 3H^2}
\,.
\end{equation}
The first two (the radius of the shell $R$ and its peculiar velocity $v$) describe the motion of $S$, the second two (the ratio of two expansion rates  ${\cal H}^-$ and the density contrast $\delta^-$)  give the boundaries conditions between the inner  and the outer regions, and the last two (the FL parameters $\Omega$ and $\lambda$) describe the background cosmological model. Herein, the initial values of the variables $v$ and ${\cal H}^-$ are given at redshift $z=z_i$ ({\it i.e.\/}, at time $t=t_i$). We choose them as follows :
\begin{equation} \label{InitParam}
z_i=100,\quad v_i=0,\quad {\rm and } \quad {\cal H}^-_{i}=1
\,.
\end{equation}
It turns out that the dependence of the expansion history of $S$ on $v_i$ and ${\cal H}^-_i$ is negligible as long as these initial data correspond to a sufficiently early epoch\footnote{Similarly as in the linear perturbation theory, once the ``decaying mode" becomes negligibly small, the eventual dynamics does not depend on initial conditions.} (typically, $z_i>10$).  In order to compare the theoretical predictions with observations, the remaining model parameters are settled to the present values at redshift $z=0$, i.e., $R_{0}$, $\delta_0^-$, $\Omega_{0}$ and $\lambda_{0}$,  by means of an iterative method\footnote{Namely, the initial values $\Omega_i$ and $\lambda_i$ are adjusted to find the present observed values ({\it e.g.\/}, $\Omega_{0}=0.3$ and $\lambda_{0}=0.7$) by solving the Friedmann equation for the outer region. We also solve the Friedmann equation for the inner region, which gives the relation between the variables of the inner region and those of the outer region by assuming the value $\delta_{0}^-$. Then, we solve the evolution equation for $R$ with time. The initial radius $R_i$ is also adjusted by means of an iterative method to obtain the present size of a void $R_{0}$.}.

In addition to the above parameters, we introduce the curvature parameter
\begin{equation}\label{curvature}
\kappa= \Omega+\lambda-1
\,,
\end{equation}
which is used to classify the  FL cosmological models. It stands for the dimensionless expression of the scalar curvature of the comoving space, which is identified to an integration constant in the Newtonian dynamics \cite{Fliche_Triay}.

Because the outer region of a void is described by the cosmological model, in order to analyze the effects of cosmological parameters on the proper dynamics of voids, we have first to study their contributions to the background cosmological dynamics. Although all cosmological parameters intervene to define the dynamics at large scale,  it is known that the universe undergoes three phases;  matter dominated(if $\Omega\neq 0$), curvature dominated (if $\kappa\neq 0$), and vacuum-energy dominated (if $\lambda\neq 0$)  expansion phases, depending on the relative magnitudes of cosmological  parameters ({\em e.g.\/} see  \cite{Triay97}).
 In the present investigation, we assume $\Lambda> 0$ and hence the Hubble parameter $H$ decreases with time and eventually toward its asymptotic value
 \begin{equation}\label{Hasymptotic}
H_{\infty}=\lim_{a\to\infty}H=\sqrt{\frac{\Lambda}{3}}
\,.
\end{equation}
In that case, there are two distinct behaviors which are characterized by the sign of $\kappa$ ( see Eq.\,(\ref{curvature})): Either $H$ decreases monotonically toward $H_{\infty}$ (if $\kappa \leq 0$) or it reaches first a minimum $H_{\rm min}<H_{\infty}$ and it eventually grows toward $H_{\infty}$ from downward (if $\kappa >0$ ). This minimum value is found to be
 \begin{equation}\label{extremum}
H_{\rm min}=H_{\infty}\sqrt{1-\frac{4\kappa_{0}^{3}}{27 \lambda_{0}
\Omega_{0}^{2}}}<H_{\infty}
\end{equation}
at $a=\frac{3}{2} \Omega_{0} \kappa_{0}^{-1}$, where $\kappa_0 $ is the present value of the dimensionless curvature constant. Near this minimum, the universe experiences a {\it loitering period\/}.

\section{Dynamics of voids}
\label{void_dynamics}

With initial conditions given in Eq.\,(\ref{InitParam}), we suspect that,  a spherical void tends to grow because the under-dense inner region expands faster than the outer region.\footnote{In the Newtonian dynamics of an empty spherical region, it is understood by the fact that each point of $S$ is radially accelerated by the attractive force of its closest denser outer region, because of the symmetry of its  shape.} To describe such a process in the Newtonian dynamics, instead of the peculiar velocity $v$, we have used the dimensionless quantity
$y=\dot R/HR$,
which enables us to disentangle/define a ``proper motion'' of the shell $S$ from the background expansion velocity. Here in order to confirm the result in the Newtonian dynamics, we discuss the same variable $y$ in GR, which reads
\begin{equation}\label{CorrectiveFactor}
y\equiv \frac{d{R}/dt}{ H{R}}=1+ \frac{f'v}{ HR}
\,.
\end{equation}
The variable $(y-1)$ works as the {\em correction factor\/} of the peculiar velocity $v$ of the shell $S$ to the Hubble expansion velocity in the outer region. 

The proper motion of a void can be found from Eq.\,(\ref{CorrectiveFactor}) in the comoving space, which frame is adapted to the investigation of large scale structures since one obtains a genuine interpretation of their dynamics\footnote{Indeed, when the co-existence of the CMB blackbody spectrum with the recession of galaxies is paradoxical with respect to Newtonian mechanics, these observations are reconciled within a description of a thermodynamical equilibrium in GR \cite{JMS74}. The set of points in the space-time from where the CMB can be observed with a black body spectrum at a given temperature $T_{0}$ defines the comoving space and provides us with a universal reference frame as long as FL model is in agreement with observations.}. With this schema in mind, we investigate the dynamics of an empty spherical region in the universe. As for the non-empty case, we shall discuss it in Appendix \ref{appendix_A}.

\begin{figure}[ht]
\begin{center}
\includegraphics[width=8cm]{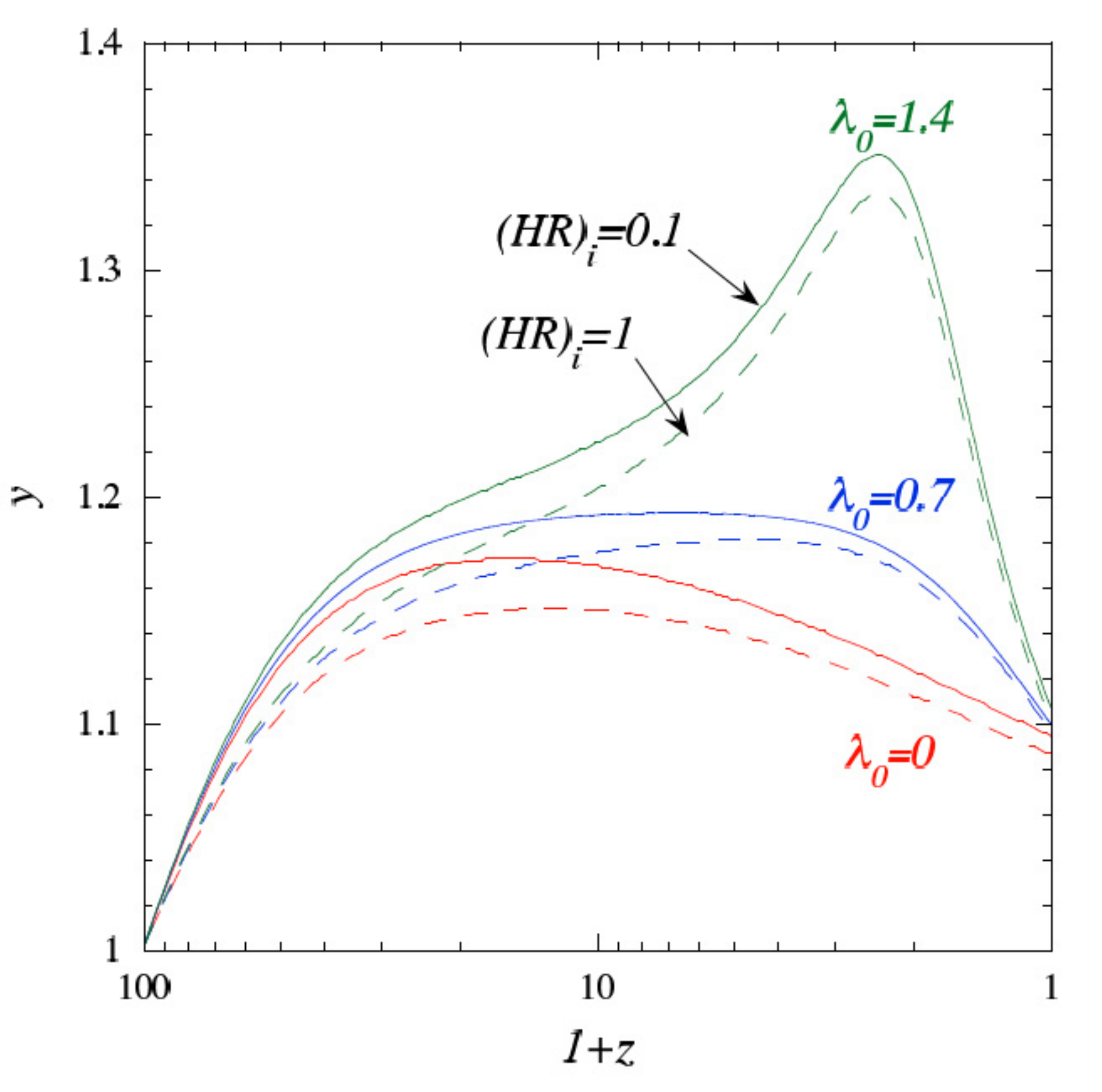}
\hskip 1.0cm
\caption{The {\em correction factor\/} to the Hubble flow --- For two voids with initial radii $R_i=0.1H_i^{-1}$ (solid line) and $H_i^{-1}$ (dashed line), the correction factor $y$ is depicted in terms of the redshift of the background universe $z$ by assuming three cosmological models defined by $\Omega_{0}=0.3$ and $\lambda_{0}\in\{0,0.7,1.4\}$.
}
\label{fig:vel_void}
\end{center}
\end{figure}

We analyze the behavior of a spherical empty void embedded in the background universe characterized by $\Omega_{0}=0.3$ and $\lambda_{0}\in\{0, 0.7, 1.4\}$, which represents distinct cosmological  models. For each one, the evolution of $y$ with time has been calculated for two spherical voids. Their radii at initial time, which we set $z_i=100$, are chosen to be 10\% and to 100\% of theinitial horizon size; {\it i.e.\/}, $R_i= \varpi H_i^{-1}$ with $\varpi\in\{0.1,1\}$. The results are described in Fig.\,\ref{fig:vel_void}, where the correction factor $y$ is given in terms of  the redshift (or the scale factor $a$). We investigate the  effect of $\Lambda$ on the void dynamics by comparing the $\lambda_{0}=0.7$ case (which will be used as a standard) with the other ones. Our analysis, which reflects the standpoint of the observer since it is based on the measurable quantities, is as follows~:
\begin{enumerate}

\item The voids always grow faster than Hubble expansion ({\it i.e.\/}, $y>1$) along their evolution. The peculiar velocity starts with a huge burst and decreases asymptotically toward Hubble velocity; the present value is higher than the Hubble flow by about $10\%$.

\item  The smaller the radius gets, the higher the peculiar velocity is. Namely, the void with the largest initial size $R_i=H_i^{-1}$  grows slightly  slower, which provides us with the minimal peculiar velocity.

\item While the dependence on $\lambda_{0}$ is not significant in the beginning, its effect eventually appears in the evolution, and the correction factor $y$ increases with $\lambda_{0}$. With $\lambda_{0}=0.7$, the peculiar velocity reaches the value higher than the Hubble expansion  at $z \sim 10$ by about $20\%$. 

\item  The redshift $z^{\star}$ when $y$ reaches its maximum $y_{\rm max}=y(z^{\star})$ decreases with $\lambda_{0}$. It corresponds to $z^{\star}\sim 40$ for $\lambda_{0}=0$ and to $z^{\star}\sim 1.7$ for $\lambda_{0}=1.4$. For the intermediate value $\lambda_{0}=0.7$,for which the corresponding curve shows a plateau,  $\Lambda$ gives the maximal contribution to $v$  at  $z^{\star}\sim 1.7$, which corresponds to  30\% of that of the matter density (when it is compared to the case of $\lambda_{0}=0$).

\item Let us pay attention to the presence of a bump on the curve, which becomes visible for $\lambda_{0}>0.7$ ({\it i.e.\/} when $k>0$). It is caused by the fact that  the universe experiences a loitering period of cosmological expansion, while the void then continues its own expansion.
 
\item The present values $y_{0}$ related to $\Omega_{0}=0.3$ but to different values for $\lambda_{0}\in\{0, 0.7, 1.4\}$ are very close to each other, while the variation of $y$ with time (or with $z$) depends undeniably on $\lambda_{0}$. In other words, the $\Lambda$ effect, which accounts for a deviation between these curves, is not substantial nowadays.

\item It is remarkable that the present GR approach for the sub-horizon void confirms the previous result found in the Newtonian dynamics \cite{Fliche_Triay}. Even for a relativistic spherical void with a horizon size radius $R_i=H_i^{-1}$, the relativistic effect turns out to be weak.

\end{enumerate}

To disentangle a $\Lambda$ effect on the dynamics of a void is quite a tricky task since all the model parameters intervene in its dynamics. It is a matter of fact that the voids expansion is magnified with $\lambda_{0}$ for a fixed value of $\Omega_{0}$. At first glance, such a feature could be understood as a gravitational local effect similar to the one described by the Schwarzschild solution of the Einstein equations with a cosmological constant $\Lambda$. Actually, this interpretation is an artifact due to the comparison of the $\Lambda$ effect between different cosmological models defined by the curvature parameter $\kappa_{0}$. The $\Lambda$ effect intervenes on the expansion of the outer region at large scale and not as being part of the own void dynamics. Such a property can be understood from a dimensional analysis of the Einstein equations, which shows that the presence of $\Lambda$ in the gravitational field is felt only at great (cosmological) distances \cite{Triay_2005}.
As a result, the excess of the peculiar velocity $v$ of an empty void with respect to the Hubble expansion velocity depends mainly on $\Omega_{0}$ \cite{SMS}, which is related to the density of the outer region. Such a behavior is similar in the case of an under-dense region but weighted by the density contrast $\delta^-_{0}$. With respect to the observer's standpoint, the dependence on $\lambda_{0}$, which intervenes on the Hubble expansion, remains extremely weak at high redshift $z>20$ and low redshift $z\sim0$, but it appears prominent at redshift $z\sim 1.7$. At this redshift, the peculiar velocity $v$  for $\lambda_{0}=0.7$ ({\it i.e.}, $\kappa_{0} =0$) is magnified by up to $50\%$ compared with the case of $\lambda_{0}=0$  ({\it i.e.}, $\kappa_{0} =-0.7$) (see Fig\,\ref{fig:vel_void}). Such features characterize $\lambda_{0}$ through the behavior of  the cosmological expansion $H$. If  $\Lambda\geq 0$ then  $H$ decreases with time and will reach its asymptotic value $H_{\infty}$. There may be a  loitering period during which a spherical void can grow much faster compared to the cosmological expansion. It clearly appears as a bump in the curve for $\lambda_{0}=1.4$ in  Fig\,\ref{fig:vel_void}.  Such a feature could be used to evaluate the value of $\lambda_{0}$by a statistical analysis of cosmological parameters from observational data \cite{Fliche_Triay}. Here we will show one simple example, by which the peculiar velocity of a void shell creates an observable deformation of a void shape in the redshift space.

\section{Redshift of a void}\label{redshift_space}

To analyze the large scale structures in the universe, one makes use of the {\it redshift-space\/}. It is identified to the direct product of the observer celestial sphere and the redshift of sources $z$. However, the images of structures are deformed because the redshift, although it is used as {\em distance indicator\/}, does not provides us  with an exact value of the distance. As a result, the image of a spherical void located at cosmic distance does not appear spherical in the redshift space\footnote{The redshift space is often used by observers because it is one of the simplest ways to construct the large scale structure at cosmological distance scale from observational data.}. The  shape of a void reflects its dynamical characteristics. Namely, the redshift of photons emitted from galaxies at the edge of the void (the shell S) takes into account its own expansion as well as the evolutionary history of the universe.\footnote{A sphere which expands in the comoving space with the cosmological expansion (but not a static one) provides us with a deformed image in the redshift space because of the evolutional effect of the universe. As we shall see later, the peculiar velocity of a void shell enhances this tendency and its deformed shape includes the information about cosmological parameters.}

Hereafter, we restrict our analysis to a spatially flat spaces of  the FLRW world model, which is characterized by a single parameter $\Omega_{0}=1-\lambda_{0}$, in addition to the Hubble constant $H_{0}$.  In Sec.\,\ref{Zdeformation},  the shape of a void in the redshift-space is modeled. We focus to an empty void ($\rho^{\rm (m)}_-=0$), in which the deformation reaches its maximal magnitude (see Appendix\,\ref{appendix_B}). A qualitative analysis is performed by  numerical  calculations in Sec.\,\ref{Nresults}.

\subsection{Redshift by proper motion of a shell}\label{Zdeformation}

The void appears as the intersection of the observer's past  light cone with the world volume $\Sigma$ of the void shell $S$. We shall visualize such an image of the intersection by means of the redshift of the source on $S$ and its angular position on the celestial sphere. With the Euclidian spatial coordinates, the FLRW metric reads
\begin{eqnarray}\label{cartesian}
ds^{2}=-dt^{2}+a^{2}(t)(dx^{2}+dy^{2}+dz^{2})
\,.
\end{eqnarray}
\begin{figure}[ht]
\begin{center}
\includegraphics[width=9cm]{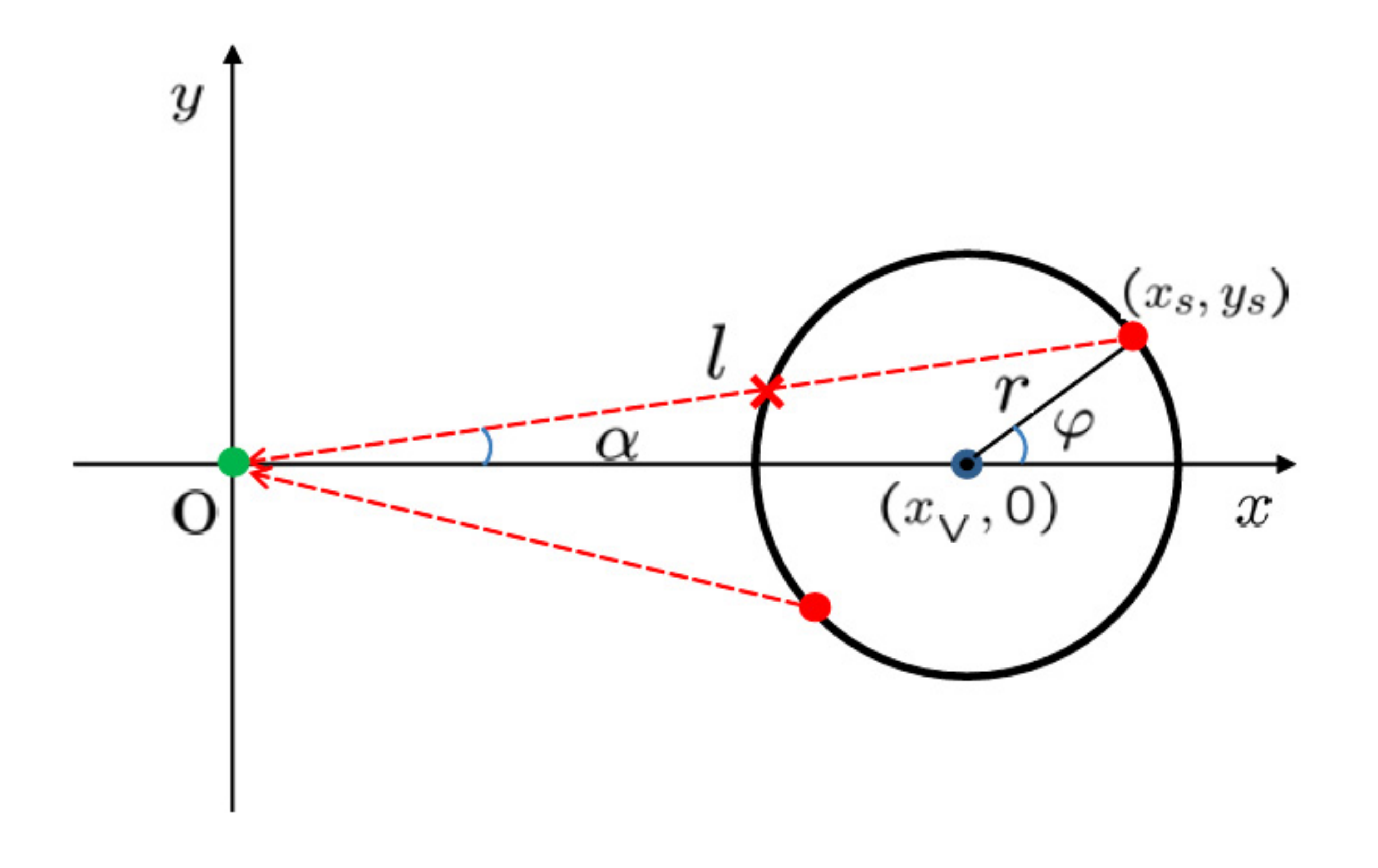}
\caption{The light rays from a void shell to the observer. --- The observer $O$ is located at the origin $(0,0)$, the $x$-axis stands for the line of sight toward the void center $(x_{\rm V},0)$. Its boundary layer (the shell $S$) is represented by the circle of comoving radius $r=r(t)$. The light ray emitted from a point $(x_s,y_s)$ located on the backside of $S$ with respect to the observer at comoving distance $l$, propagates inside  the void before crossing $S$ at a point $\times$, and then reaches the observer
 $O$; $\varphi$ and $\alpha$ are the viewing angles of the light source with respect to the $x$-axis from the observer $O$ and from the center of $S$,
 respectively. The light ray from the foreside point $(x_s',y_s')$ propagates directly toward the observer $O$.}
\label{fig:void_parameter}
\end{center}
\end{figure}
From the symmetry, without loss of generality, it is enough to analyze the light rays propagating  on a two-dimensional plane,  which we choose the $xy$-plane (see Fig.\,\ref{fig:void_parameter}). The observer $O$ is  located at the origin of the coordinates system and the center  of the void shell lies at $x=x_{\rm V}$ on the $x$-axis.  The comoving radius of the void is given by  $r(t)=f(\chi^{(\Sigma)}(t))$,  the  peculiar expansion velocity is $v=a{dr/dt}$, and the Lorentz factor is $\gamma=1/\sqrt{1-v^{2}}$. 
A point $X_sS$ on $S$ is described by its spatial (comoving) coordinates  as $X_s\equiv(x_s,y_s,0)$, where
\begin{equation}\label{shell}
 x_s=x_{{\rm V}}+r\cos\varphi,~y_s=r\sin\varphi
\,,
\end{equation}
and the 4-velocity of $S$ at $X_s$ reads 
\begin{equation}\label{Vshell}
u_s^{\mu}=
\left(\gamma,~{\gamma v\over a}\cos\varphi,~{\gamma v\over a}\sin\varphi,~0
\right)
\,.
\end{equation}
The 4--momentum of a light ray reads
\begin{eqnarray}\label{k0}
k^\mu=E\left(1,-{\cos\alpha\over a},-{\sin\alpha\over a},0\right)
\,.
\end{eqnarray}
where $E$ is the photon energy and $\alpha$ is the angle between the light ray  and the $x$-axis (see Fig.\,\ref{fig:void_parameter}). 
The 4-velocity of the observer reads  $u^{\mu}_o=(1,~0,~0,~0)$. The redshift of a photon emitted from $X_s$ and  received by the observer is then given by 
\begin{equation}\label{z}
1+z_{s}={(u^{\mu}k_{\mu})_s\over(u^{\nu}k_{\nu})_o}={E_s\over E_o}
\gamma\left[1+v\cos(\varphi-\alpha)\right]
\end{equation}
As we shall see later, the geometrical effects inside the void are negligible.  Hence, instead of solving the geodesic equation to calculate $E_s/E_o$,  we use the (homogeneous) approximation
\begin{equation}
{E_s\over E_o}={E_s\over E_o}\bigg|_{\FLRW}:=
{a_o\over a_s}=1+z\,,
\end{equation}
where $z$ stands for the (cosmological) redshift when the photon is emitted. This  is free from  the Doppler effect due to the shell expansion. Eq.\,(\ref{z})  transforms
\begin{equation}\label{zz}
1+z_{s}=\left(1+z\right)\gamma\left[1+v\cos(\varphi-\alpha)\right]
\end{equation}
The comoving coordinates of the emission event read
\begin{equation}\label{lightpath}
x_s=l(z)\cos\alpha,\quad y_s=l(z)\sin\alpha,\quad
l(z)=\int^z_{0}{dz\over a_{0}H(z)}
\end{equation}
where $l(z)$ is the comoving distance from the observer to $X_s$. Hence, for a given $\alpha$, from Eqs.\,(\ref{shell}) and (\ref{lightpath}), we find the redshift $z$ when 
the photon is emitted and two intersecting points (two values of $\varphi$ that correspond to points located in the foreside and in the backside of $S$ with respect to the observer). Since $v$ is derived from the dynamics equations, we obtain $z_s$ in terms of $\alpha$, which relation provides us with the shape of a void in the redshift space. Its numerical procedure is summarized as follows:
\begin{enumerate}
\item[(i)] Solve our basic equation for the void dynamics to find $r=r(z)$ 
and $v=v(z)$.
\item[(ii)] For a given angles $\alpha$, find the intersecting points
 ($\varphi$)
of the light path  with the shell and the time $z$ when the photon
 is emitted.
\item[(iii)] Calculate the redshift $z_{s}$ of the void shell for each angle 
$\alpha$.
\item[(iv)] Draw the shape, i.e. $(z_x,z_y)=(z_{s}(\alpha)\cos\alpha,~z_{s}
(\alpha)\sin\alpha)$.
\end{enumerate}

The order of magnitude of the contribution to redshift $z_{s}$ from the peculiar velocity $v$ is estimated by the related Doppler shift as\footnote{It must be emphasized that the interpretation of the redshift $z$ due to the Friedmann expansion as the Doppler shift is quite tricky without a model which enables us to disentangle the Doppler effect
by the motion of an object  from the gravitational redshift.}
\begin{equation}
\Delta z_{s}|_{{\rm Doppler}}\approx v\cos(\varphi-\alpha)\approx0.1HR 
\cos(\varphi-\alpha),\qquad 
{\rm for } \quad\Omega=0.3
\,.
\label{Doppler}
\end{equation}

\subsection{Other effects on the redshift of a void}\label{Dinhomogeneity}
In the previous subsection, we have evaluated the redshift of a void just due to the Doppler effect by the proper motion of the shell. However, there are two other effects which contribute to the redshift. Those are the following two effects caused  by density inhomogeneities.

\begin{itemize}
\item The Sachs-Wolfe effect. --- The Sachs-Wolfe effect vanishes \cite{SSY} because there is no potential difference between the shell $S$ and the background universe under the thin-shell approximation. The non-vanishing redshift is found for  the integrated Sachs-Wolfe (ISW) effect \cite{TV}, which is evaluated by
\begin{equation}\label{ISW}
\Delta z_{s}|_{{\rm ISW}}\sim(HR)^3
\end{equation}

\item A gravitational lens effect. --- For the light ray from the backside of a void, the light is bended at some point on the foreside shell 
(the mark `$\times$'' in Fig.\,\ref{fig:void_parameter}).
The scattering angle at the shell (see Appendix\,\ref{appendix_B}) 
is approximately given by 
\begin{equation}\label{Deltaalpha}
\Delta\alpha\approx0.25HR\sin(\varphi-\alpha),\qquad
{\rm for }\quad\Omega_0=0.3
\end{equation}
Note that $\Delta\alpha=0$ for a photon which goes through the center 
($\varphi=\alpha=0$) of the void or for one which
comes from the edge ($\varphi-\alpha=\pi/2$).
Hence the ratio of the maximum radius to the minimum one 
is not modified, although 
this light bending may slightly change the shape of a void
 in the redshift space.

In the present thin shell void model, 
the only relevant quantity is related to the change of the optical path length
 inside the void caused by a deflection, which is $\sim R(\Delta\alpha)^2$.
 Hence, the redshift due to the gravitational lens effect reads
\begin{equation}\label{GW}
\Delta z_{s}|_{{\rm GL}}\approx{R(\Delta\alpha)^2\over ax_{{\rm V}}}
\approx0.06{r\over x_{{\rm V}}}(HR)^2\sin(\varphi-\alpha),\qquad
{\rm for }\quad\Omega_0=0.3
\end{equation}

\end{itemize}

As a result, the contribution to redshift of these inhomogeneity effects 
turned out to be negligible compared to that of the Doppler effect given 
in Eq.\,(\ref{Doppler}) for voids with sub-horizon sizes.

\subsection{Numerical results}\label{Nresults}
For the analysis of a spherical void in the redshift space, we compare
 the image of its shell $S$ with the one of a comoving sphere of constant 
comoving radius $r$, which we use as a reference shape. 
In the following figures, 
the image of $S$ is represented by the intersection 
of $\Sigma$ in the redshift space 
with the light cone of the observer. 
The event in the redshift space is 
described by the coordinates ($z\cos{\alpha}$,$z\sin{\alpha}$) 
in the two-dimensional $xy$ plane.
 As we shall see from our numerical results, 
the image of $S$ (red dots) looks roughly like an ellipse whose center is
 located at the redshift $z_{\rm V}$. 
Its major axis (of length $2\Delta z_{||}$)
 coincides with the line of sight of the observer, and the minor axis 
(of length $2\Delta z_{\perp}$) is free from the elongation.
As a result, the shape of a void in the redshift space is elongated 
in the direction of the line of sight.\footnote{It is the
 opposite deformation of Kaiser effect, since the peculiar velocities are 
oriented outward.} A parameter to quantify such an effect is the 
{\it deformation ratio\/}\footnote{In the present case, it is given by 
the ratio of the semi-major axis to the semi-minor axis of the approximate 
ellipse since $\Delta z_{||}>\Delta z_{\perp}$. It is more appropriated than
 the eccentricity for measuring the elongation of the image 
because it shows whether $S$ is expanding (${\cal R}>1$) or contracting
 (${\cal R}<1$).}
\begin{equation}\label{DefRate}
{\cal R}={\Delta z_{||}\over \Delta z_{\perp}}
\,.
\end{equation}
The shape of the reference void
 is also an ellipse (black dots), which is due to 
the evolutionary effect of the background universe. 
It shares the minor axis with the image of $S$ 
because there is no
redshit due to the Doppler effect in that direction,
and shows a smaller deformation ratio ({\it i.e.\/}, it is less elongated).
 The comparison between these images provides us with an estimate 
on the magnitude of the peculiar  velocity of the void shell (relatively to 
the Hubble flow). 
\begin{figure}[ht]
\begin{center}
\includegraphics[width= 7cm]{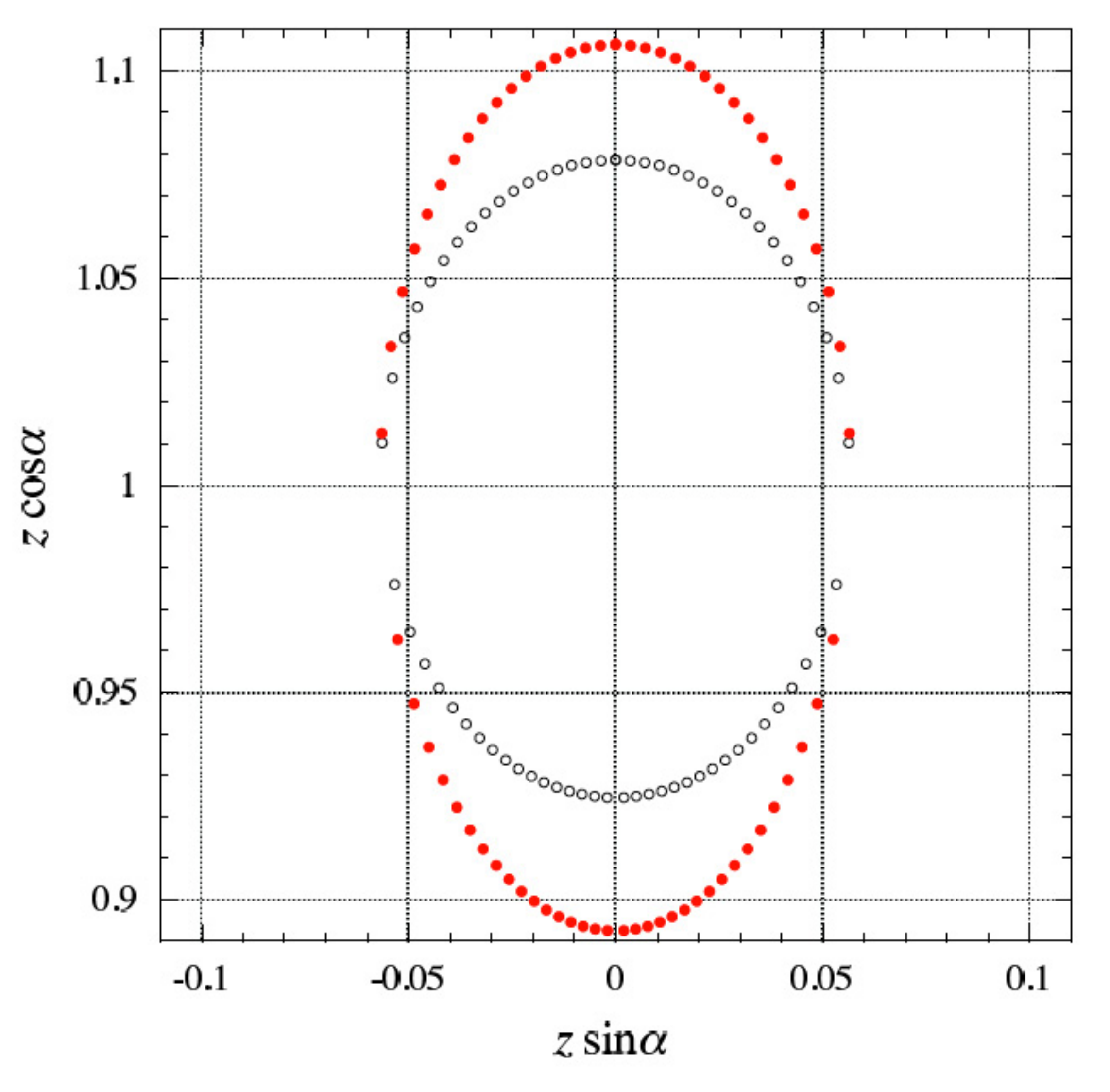}
\hskip 1cm
\includegraphics[width=7cm]{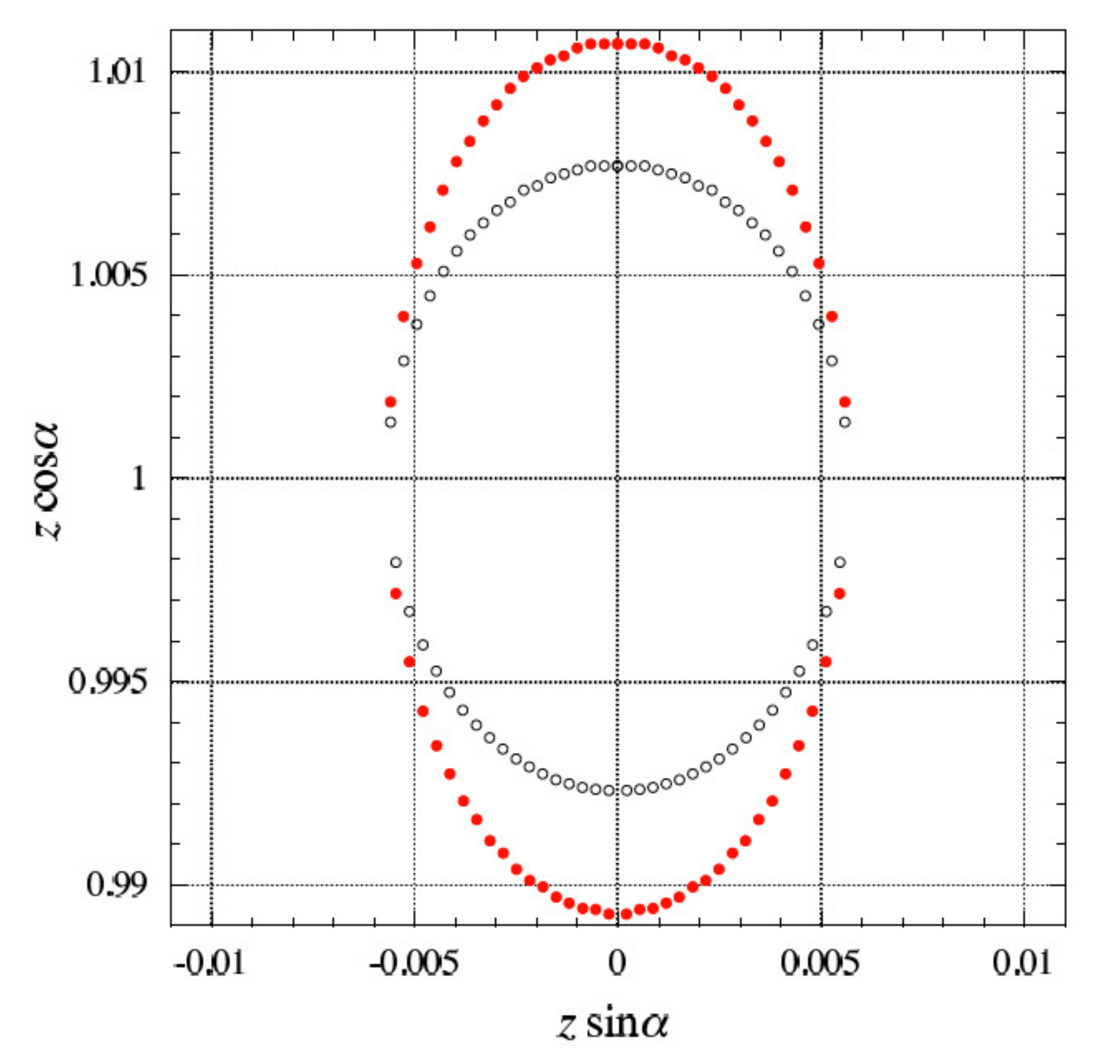}
\\
(a) $R_{0}=0.1H_{0}^{-1},\quad z_{\rm V}=1$ \hskip 2cm (b) 
$R_{0}=0.01H_{0}^{-1},\quad z_{\rm V}=1$
\caption{Empty spherical voids in the redshift space. ---
 The images (red dots) of their boundary layers ({\it i.e.\/}, 
the void shell $S$)
 are depicted for the case of $\Omega_{0}=0.3$ and $\lambda_{0}=0.7$. 
The present values of their radii are (a)~$R_{0}=0.1H_{0}^{-1}$ and 
(b)~$R_{0}=0.01H_{0}^{-1}$. The observed radii at $z_{\rm V}=1$ are~: 
(a)~$R_{\rm V}=0.0437H_{0}^{-1}$ and (b)~$R_{\rm V}=0.00437H_{0}^{-1}$. 
The images of the standard static spheres (small black dots)
are also displayed.}
\label{fig:zspace1}
\end{center}
\end{figure}

In addition to the above main 
results, we analyze the  void shapes with respect to the following
observational viewpoints~:\\[.5em]
(1) Dependence of properties of voids 
\begin{itemize}

\item Size ---
For the case of $\Omega_0=0.3$, we have analyzed 
the images of voids located at $z_{\rm V}=1$ 
with two different present radii
$R_{0} \in\{0.1H_{0}^{-1},\,0.01H_{0}^{-1}\}$,\footnote{
 The observed void size is
\begin{equation}
R_{\rm V}(z_{\rm V})\equiv a_{0}r(z_{\rm V})=
R_{0}\times {r(z_{\rm V})\over r(0)}<R_{0}
\,,
\end{equation}
because the photons were emitted at $z=z_{\rm V}$ (the past).} 
which are depicted in Fig.\,\ref{fig:zspace1}.
The image of the void with the present size 
of $R_{0}=0.1H_{0}^{-1}$ shows a small 
fore-back asymmetry (see Fig.\,\ref{fig:zspace1}\,(a)). Such a feature can be
 interpreted as an {\em evolutionary effect\/} due to the photons emitted 
 at different times
from points of the expanding shell $S$.
 It does not substantially affect the shape.
It is not visible on 
the image of the void with a smaller radius $R_{0}\sim0.01H_{0}^{-1}$, 
for which 
it is expected to be weaker because the time difference is small 
(see Fig.\,\ref{fig:zspace1}\,(b)). 
Note that these shapes are similar while the radii of 
corresponding voids differ 
by one order of magnitude. Therefore, the deformation ratio ${\cal R}$ turns
 out to be (almost) independent of the size of a void, for which  the elongation 
in the redshift space is proportional to $\Delta z_{s}\sim v\sim0.1HR$. 
Such an interesting feature should be  taken into account in the observation 
of voids.
\item Distance ---
\begin{figure}[ht]
\begin{center}
\includegraphics[width= 7cm]{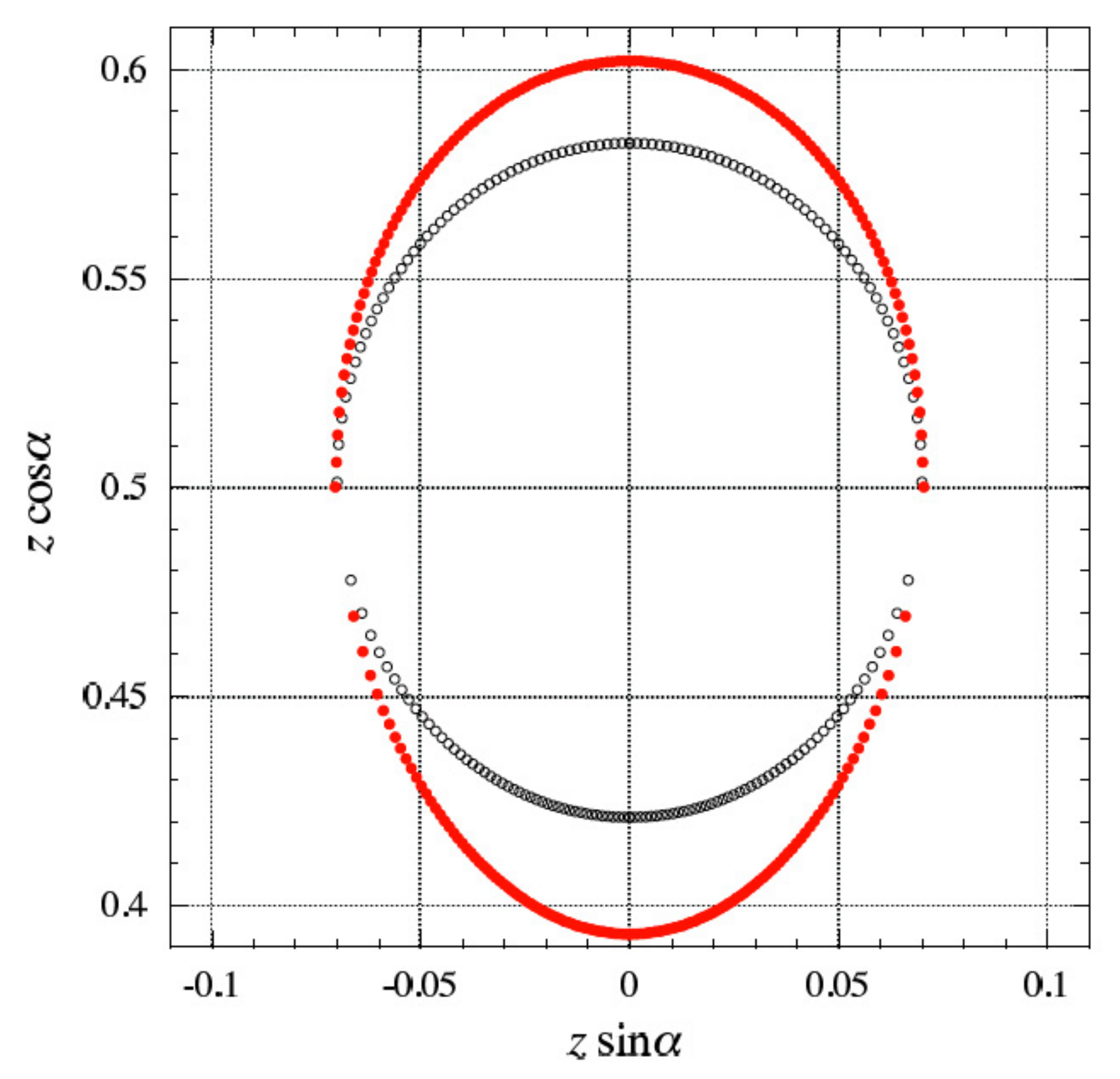}
\hskip 1cm
\includegraphics[width= 7cm]{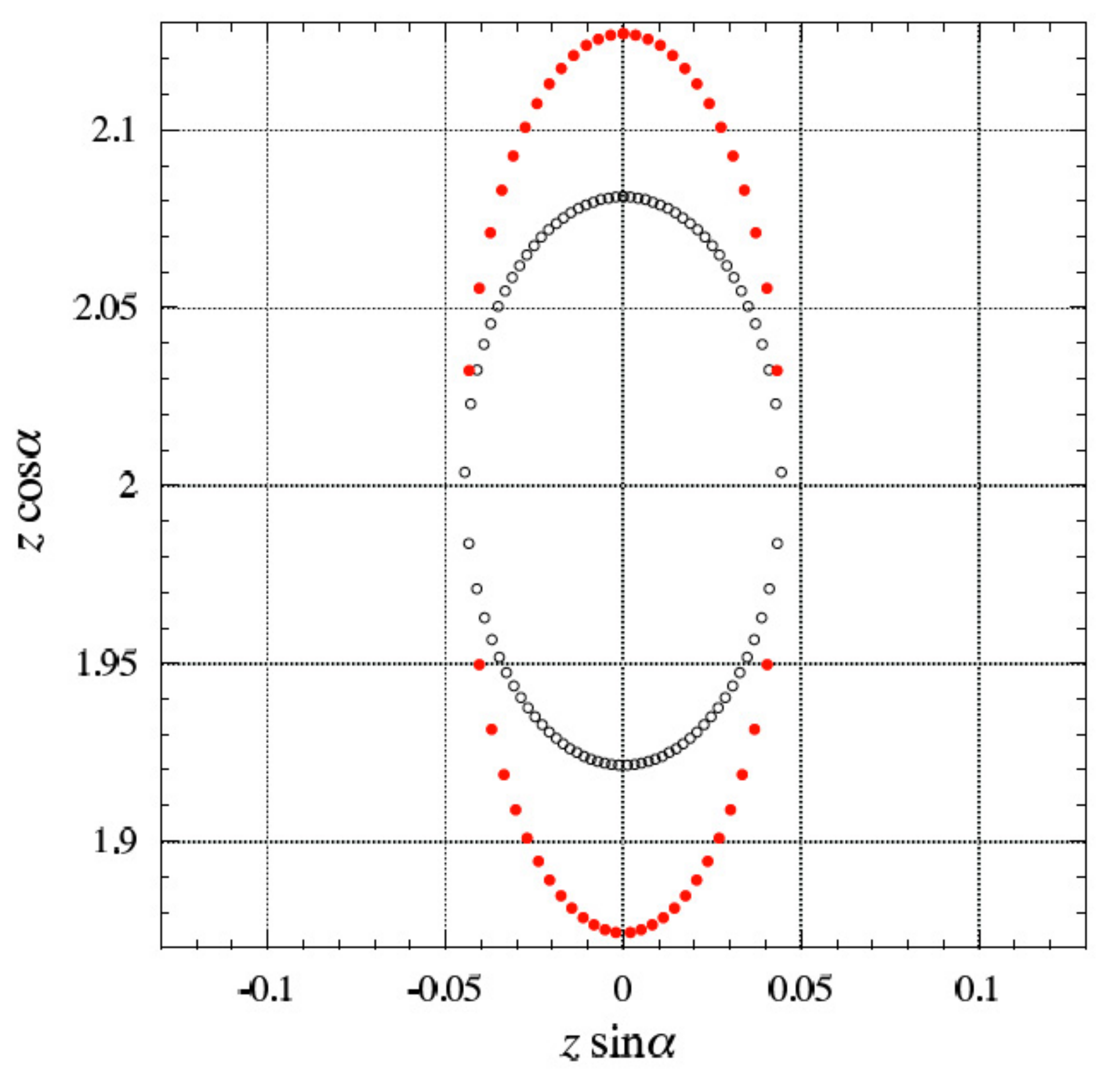}
\\
(a) $R_{0}=0.1H_{0}^{-1},\quad z_{\rm V}=0.5$, \hskip 2cm (b) 
$R_{0}=0.1H_{0}^{-1},\quad z_{\rm V}=2$
\caption{The images of empty spherical voids 
with the radius $R_{0}=0.1H_{0}^{-1}$ 
in the redshift space located at two different redshifts  
(a) $z_{\rm V}=0.5$ and (b)~$z_{\rm V}=2$.
The observed radii of voids are (a) $R_{\rm V}=0.0616H_{0}^{-1}$ and 
(b) $R_{\rm V}= 0.0269H_{0}^{-1}$, respectively. The small 
black dots represent the images of the standard static spheres}
\label{fig:zspace2}
\end{center}
\end{figure}
For  $\Omega_{0}=0.3$, 
the images of voids with the resent radius $R_{0}=
0.1H_{0}^{-1}$ but located at two different redshift 
$z_{\rm V} \in\{0.5, 2\}$ are depicted in Fig.\,\ref{fig:zspace2}.
We find that the deformation ratio ${\cal R}$ increases 
with redshift. Such an effect is clearly seen  in Fig.\,\ref{fig:zspace3}\,(a),
in which 
 four images of voids with $R_{0}=0.1H_{0}^{-1}$ at redshift $z_{\rm V}
 \in\{0.5, 1, 1.5, 2\}$ are displayed. The reason why this elongation of
the shape 
 with the redshift (distance) $z_{\rm V}$ increases is twofold,
i.e., in the redshift space, $\Delta z_\perp$ gets 
smaller because it corresponds to the observed size $R_{\rm V}(z_{\rm V})$,
 while $\Delta z_{||}$ increases with the peculiar velocity 
of the void shell (see footnote 11).
\end{itemize}
(2) Dependence of the background cosmological models
\begin{itemize}
\item  Matter density ---
\begin{figure}[ht]
\begin{center}
\includegraphics[width=7cm]{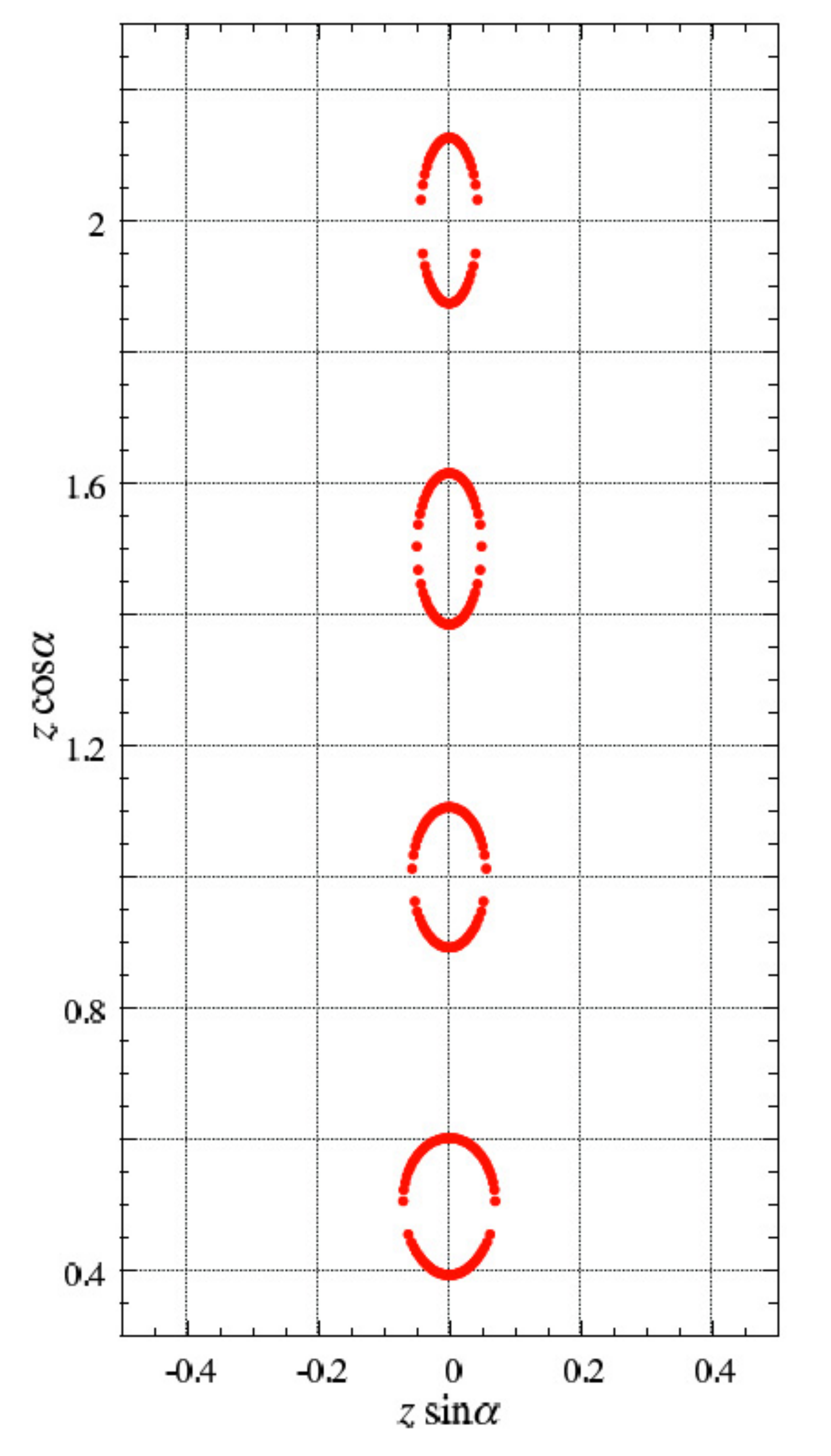}
\hskip 1cm
\includegraphics[width=7cm]{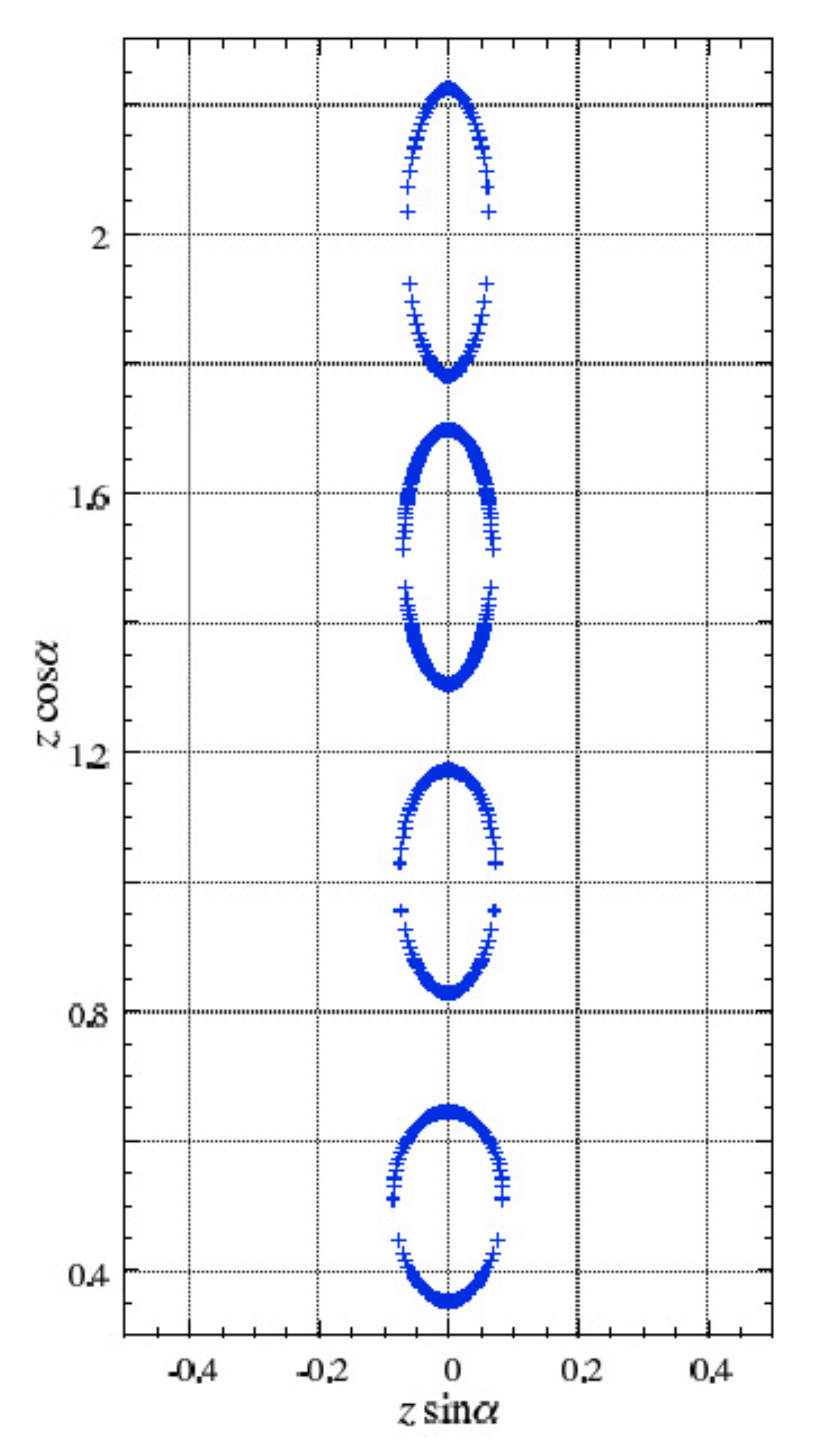}
\\
(a) $\Omega_{0}=0.3$ \hskip 3cm (b) $\Omega_{0}=1$
\caption{
The images  in the redshift space of an empty spherical void with present 
radius $R_{0}=0.1H_{0}^{-1}$ at redshift $z_{\rm V} \in\{0.5, 1, 1.5, 2\}$
 for (a)~$\Omega_{0}=0.3$ and (b)$\Omega_{0}=1$.}
\label{fig:zspace3}
\end{center}
\end{figure}
Four images of voids with $R_{0}=0.1H_{0}^{-1}$ located at redshift 
$z_{\rm V} \in\{0.5, 1, 1.5, 2\}$ are displayed in Fig.\,\ref{fig:zspace3} 
for two different values of $\Omega_{0}\in\{0.3, 1\}$.\\
By comparing these images in Fig.\,\ref{fig:zspace3}\,(a,b), 
it is clear that the deformation ratio ${\cal R}$ increases with $\Omega_{0}$
for all redshift.  Let us remind that the background matter density, which 
characterizes the void as well as the outer space where it is embedded, 
is represented through the cosmological parameter $\Omega_{0}$.

\item Deformation ratio ---
\begin{figure}[ht]
\begin{center}
\includegraphics[width=7cm]{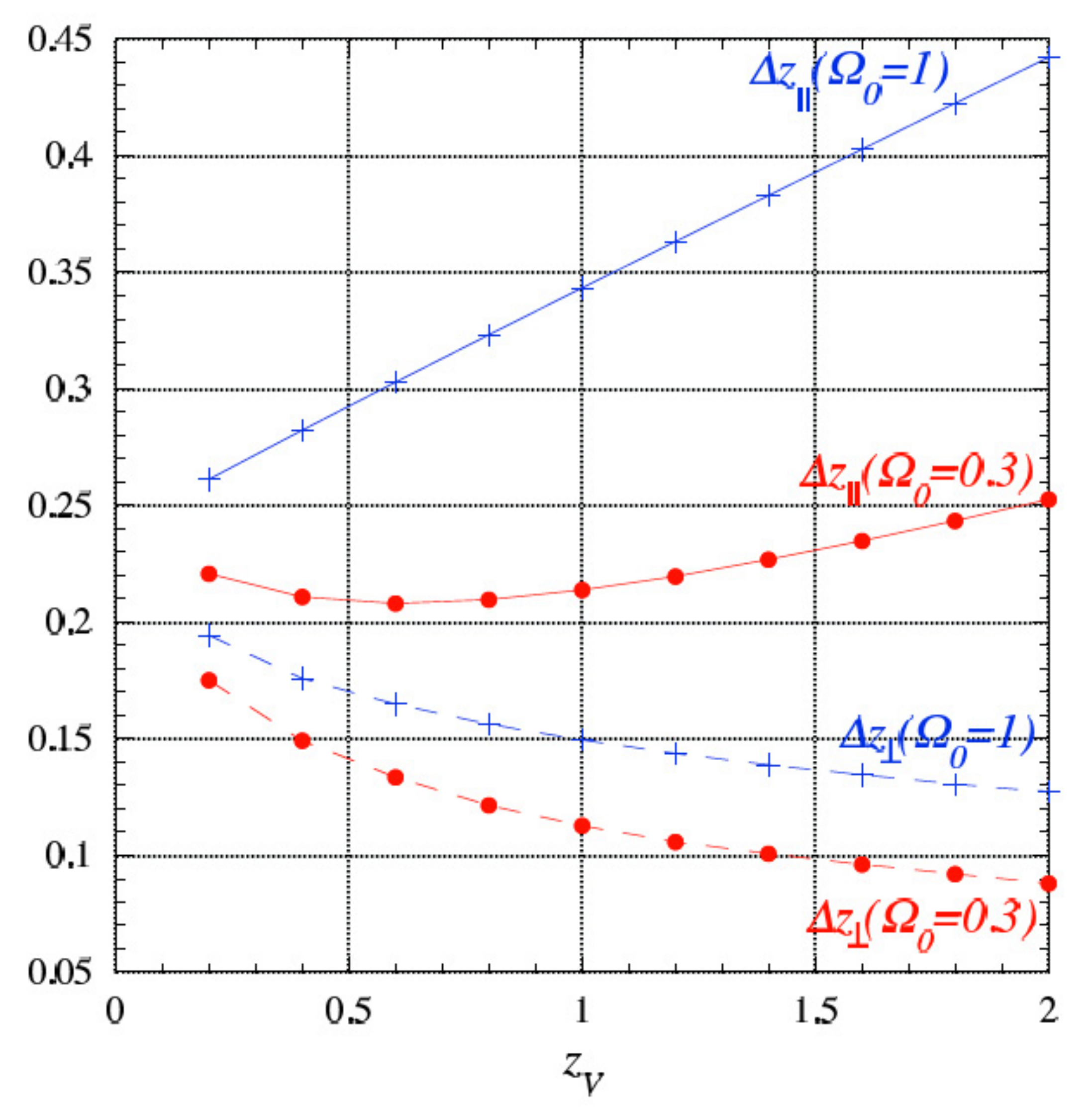}
\hskip 1cm
\includegraphics[width=7cm]{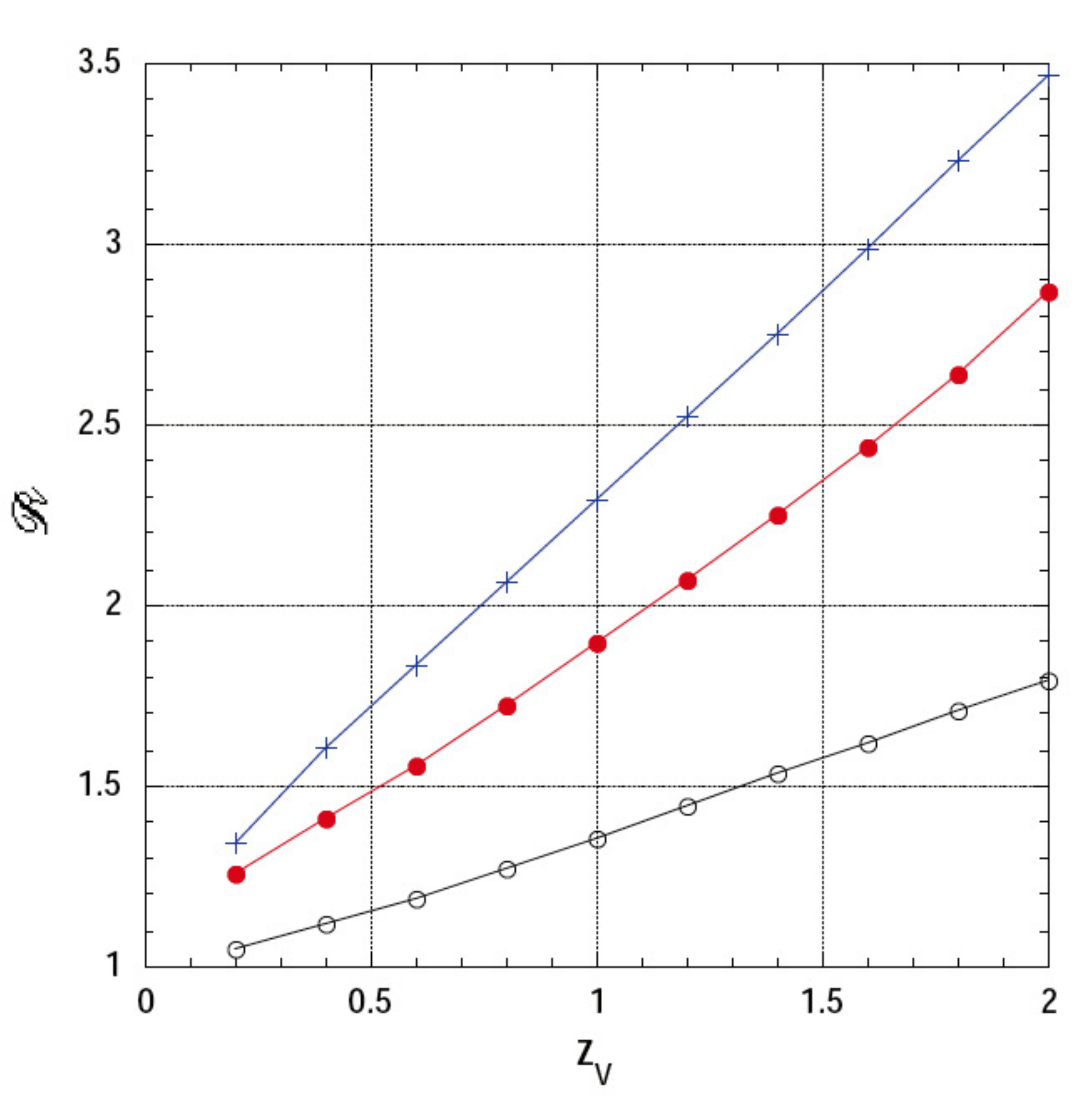}
\\
(a) $\Delta z_{\perp}$ and $\Delta z_{||}$ \hskip 3.5cm 
(b) The deformation ratio ${\cal R}$
\caption{(a)The semi-major axis $\Delta z_{||}$ and the semi-minor axis 
$\Delta z_{\perp}$ in terms of the distance $z_{\rm V}$. We find that 
$\Delta z_{\perp}$ always decreases as $z_{\rm V}$ increases because 
the observed void size gets small. On the other hand, $\Delta z_{||}$ 
increases as $z_{\rm V}$ increases due to the Doppler effect except for 
nearby voids for $\Omega_{0}=0.3$. 
(b)~The deformation ratio ${\cal R}$ of the shape of a void in the red shift 
space, which is defined by ${\cal R}=\Delta z_{||}/\Delta z_{\perp}$, where
 $\Delta z_{||}$ and $\Delta z_{\perp}$ are the semi-major axis and 
semi-minor axis radii of the void shape in the redshift space, respectively.
 The ratio ${\cal R}$ increases as the distance $z_{\rm V}$ increases. 
We also find the ratio ${\cal R}$ highly depends on the background cosmological
 model, which may give us some information on the cosmological constant. }
\label{fig:ratio}
\end{center}
\end{figure}
The property of the deformation ratio ${\cal R}$ of being independent 
on the void size motivates us to analyze this parameter with more care. 
In Fig.\,\ref{fig:ratio}(a), we confirm that $\Delta z_\perp$ is 
a monotonically decreasing function of $z_{\rm V}$. On the other hand, 
$\Delta z_{||}$ increases for $\Omega_{0}=1$ while it first decreases toward 
a minimum and eventually increases in the case of $\Omega_{0}=0.3$. 
The deformation ratio ${\cal R}$ is an monotonically increasing
 function of $z_{\rm V}$ because  $\Omega$,
 and hence the peculiar velocity of the void shell ($v \sim 0.2HR$),
  increases with  $z_{\rm V}$.
This tendency is enhanced as $\Omega_0$ increases 
(see Fig.\,\ref{fig:ratio}\,(b)). Since we focus 
on a flat universe model ($\kappa =0$), any information for $\Omega$ 
provides us with an estimate of a cosmological constant.
\end{itemize}


\section{Conclusion and Remarks}\label{Conclusion}

With the aim to understand the dynamics of voids in the universe and 
the effect of a cosmological constant $\Lambda$ on it, we analyze
 a relativistic model of a spherical under-dense region 
in a FL universe.
The inner and the outer regions 
are described by space-times with uniformly distributed dust.
 We study the effects 
of cosmological parameters on the dynamics of voids, focusing 
on the case of an empty void. 
We have calculated the peculiar  velocity of a void shell, with which  
the shell moves in the comoving space 
because the lower density region expands
faster than the outside Hubble flow. 
We show that the dynamics depends
 mainly on the density parameter $\Omega_{0}$. 
Our analysis confirms the previous
 result obtained in the Newtonian dynamics, 
namely, the relativistic effects are rather weak. 
The shape of voids in the redshift space has also been investigated
 in the case of a spatially flat cosmological model. 
It is characterized mainly by the void 
kinematics, i.e., the Doppler shift due 
to the proper motion of a void shell.
The other effects on the shape, such as the ISW effect 
and the gravitational lens effect, turn out to be rather small 
for a void with a sub-horizon size.
The void in the redshift space appears as  
an ellipse shape elongated in the direction of 
the line of sight (opposite to 
a deformation by the Kaiser effect), 
although a small modification from the ellipse shape by
an evolutionary effect may become 
perceptible with increasing the size.

Since the void dynamics depends on a background cosmological model,
such observational data  could be used to improve
 statistics of cosmological parameters, especially
that of a cosmological constant by means of the structures 
at redshift $z\approx 2$. 
It is, however, clear that our model of a single spherical void 
is too simple and it must be improved 
in order to account for shape irregularities 
due to interactions of galaxies on the edges of voids. 
In order to perform such improvements, 
one may need numerical approach.
Since the dynamics of such a void-network is much
 more complex than that of a single isolated void,
$N$-body simulation would be required.
In the conventional $N$-body simulation,
we usually assume that the space-time is described by a uniform
and isotropic universe. The expansion velocity of the background 
universe is uniform and it depends on
 the mean density of matter fluid but 
does not on the local density. 
Such an approach is justified if the density perturbations
are small.  However, if the perturbations are non-linear, i.e.,
if the density contrast is very large just as in the late stage of 
structure formation, we may have to worry about the position dependent
 background expansion just as our present void model.
In fact, according to our model, which describes the dynamics of a spherical 
under-dense region in a FL universe, such an under-dense
 region  expands faster
 than the background Hubble flow, i.e.,
the space-time region with under-dense matter distribution
 expands faster than the over-dense region.
This fact will change the dynamics of structures just as our simple case.
Hence, with the aim to understand the mechanism of 
the large scale structure formation, 
we have to improve the Newtonian $N$-body simulation. 
For this purpose, the Newton-Friedmann model 
\cite{Fliche_Triay}, which is 
based on the covariant formulation of Euler-Poisson 
equations \cite{Souriau70},
 appears to be more adapted for being implemented 
in the numerical codes, although we may need 
further study in its practical formulation.

It is interesting to note that such an investigation can be extended to test 
other gravity theories \cite{SM} or approaches for dark energy.
It is widely discussed as an alternative interpretation 
of the observed acceleration of the cosmological expansion 
\cite{dark_energy1,dark_energy2,dark_energy3,dark_energy4}, 
although a cosmological constant remains the most appropriated explanation 
according to Occam's razor.

~\\
{\bf Acknowledgements:}\\[1em]
We thank H.H. Fliche for valuable comments and discussions. KM would 
acknowledge hospitality of Centre de Physique Th\'eorique (CNRS) and 
The Universit\'e de Provence, during his stay in 2009 and 2010. This work 
was partially supported by the Grant-in-Aids for Scientific Research Fund of
 the JSPS (No.22540291) and for Scientific Research on Innovative Areas 
No.\ 22111502, and by the Waseda University Grants for Special Research 
Projects.

\appendix

\section{Spherical  under-dense void}
\label{appendix_A}
 
In this appendix, we extend our analysis of the void dynamics to
the case of non-empty void.
We assume that the inner region of a void is filled by 
a uniform distribution of dust with a density $\rho_{-}^{\rm (m)}
(<\rho^{\rm (m)})$.
The density contrast is characterized by 
$\delta^-=\rho_{-}^{\rm (m)}/\rho^{\rm (m)}-1$
($-1<\delta^-<0$).
We restrict our analysis to the case of a flat background universe 
($\kappa =0$). With appropriate initial values for the model 
parameters, we have integrated 
the evolution equations for an under-dense void, which variables are
 $(y, \Omega, \delta^-)$, until the redshift $z_0$.
If $z_0=0$, the calculation gives the present values.
It turns out, however, that the relation 
between our variables $(y, \Omega, \delta^-)$
 does not depend on  $z_0$ as well as 
the initial time $z_i$ as long as the 
``decaying mode" becomes negligibly small during the evolution.

\begin{figure}[ht]
\begin{center}
\includegraphics[width= 8cm]{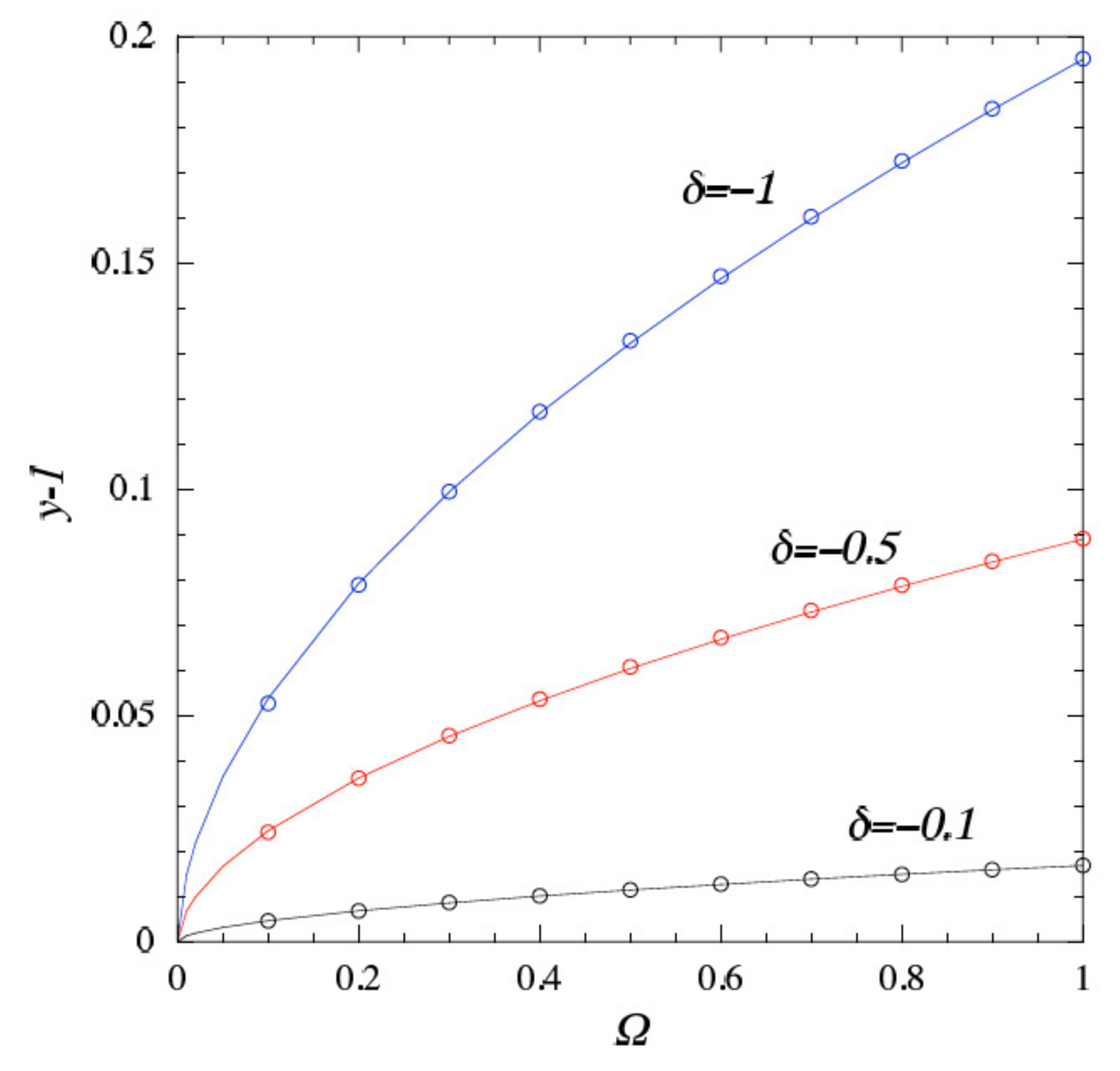}
\\
Velocity $y$ versus $\Omega$ and $\delta^-$.
\hskip 1.0cm~
\caption{The velocity $y$ of a non-empty void. --- We assume $\Omega+\lambda=1$ and $\delta^-\in\{01,-0.5,-0.1\}$. The circles and the continuous lines correspond respectively to our numerical results and to the fitting formula given in Eq.\,(3.2).}
\label{fig:vel_void2}
\end{center}
\end{figure}

The relation found by the numerical calculation
 is shown by circles in Fig.\,\ref{fig:vel_void2}.
We can find a fitting formula for the relation as
\begin{equation}
y-1={v\over HR}
={\Omega^{0.56}\over6}(|\delta^-|+0.1 |\delta^-|^2+0.07|\delta^-|^3),\quad
{\rm for} \quad \Omega+\lambda=1
\label{formula}
\end{equation}
which is shown by the solid curved in Fig.\,\ref{fig:vel_void2}. 
It approximates our 
numerical results very well.
The difference between the numerical results and the formula 
(\ref{formula}) is less than one percent.

The expansion of an under-dense region differs from that of an empty void 
($\delta^-=-1$), although the dynamics is similar to that of an empty void.
As we see from Fig.\, \ref{fig:vel_void2}, 
for a given $\Omega$, the  higher 
the inner density is ({\it i.e.\/}, $\delta^-\to 0$), 
the smaller the peculiar  velocity of the void shell (i.e., $(y-1)$) is.
This is because the space with under-dense matter expands faster than
that with over-dense matter. 

As we have seen in Fig.\,\ref{fig:vel_void}, the peculiar velocity
increases monotonically 
as  $\Omega$ increases.
The similar feature is seen for non-empty void
(see Fig.\,\ref{fig:vel_void2}).
In fact the curves in Fig.\,\ref{fig:vel_void2}
are found by extrapolating 
the curve in Fig.\,\ref{fig:vel_void} 
with (reasonable) continuous 
transformations up to the constant function $y=1$ for $\delta^-=0$. 
As the result, we find that 
the correction factor $y$ depends strongly on $\delta^-$
 as well as on $\Omega$.

\section{Deflection of the backside light ray at the boundary shell}
\label{appendix_B}

In this appendix, we derive the deflection angle for a light ray that is 
emitted from a point located on the backside of a void with respect to 
the observer. As described in Fig.\,\ref{fig:void_parameter}, it propagates 
inside the void, crosses the boundary shell $S$ at a point denoted $\times$,
and  then reaches the observer  $O$. For convenience, 
we perform the coordinate transformations of the 4-momentum $k^{\mu}$ 
in the vicinity of the shell four times as follows~:
\begin{enumerate}
\item We translate Eq.\,(\ref{k0}) from Cartesian coordinates (\ref{cartesian})
 to spherical ones (\ref{FLRW}), i.e.,
\begin{equation}
k^t=E,\quad
k^\chi=-{E\over a}\cos(\varphi-\alpha),\quad
k^{\varphi}={E\over af}\sin(\varphi-\alpha),\quad
k^{\theta}=0
\,,
\end{equation}
\item We adopt a Gaussian normal coordinate system 
$(\tau, n, \theta, \varphi)$ which covers both the outer and 
the inner parts of $S$. $n$ is the spatial coordinate normal to 
the hypersurface of the shell $\Sigma$, which corresponds to  $n=0$,
and the  3-metric of $\Sigma$ is given by Eq.\,(\ref{3metric}). 
The coordinates transformation of $k^{\mu}$ at $n=0$ from the outer FLRW frame 
to the Gaussian normal frame  gives
\begin{equation}
k^{\tau}=E\gamma[1+v\cos(\varphi-\alpha)],\quad
k^n=-E\gamma[v+\cos(\varphi-\alpha)],\quad
k^{\varphi}=-{E\over R}\sin(\varphi-\alpha),\quad
k^{\theta}=0
\,,
\end{equation}
where we have used the relation (see \cite{SM}),
\begin{equation}
{\partial t\over\partial\tau}=\gamma,\quad
{\partial r\over\partial\tau}={\gamma v\over a},\quad
{\partial t\over\partial n}=\gamma v,\quad
{\partial r\over\partial n}={\gamma\over a}
\,.
\end{equation}
\item The transformation of $k^{\mu}$ from the Gaussian normal frame to the 
inner FLRW frame gives 
\begin{eqnarray}
k^{t_-}&=&E\gamma_+\gamma_-[1+(v_+-v_-)\cos(\varphi-\alpha)-v_+v_-], \nonumber
 \\
k^{\chi_-}&=&{E\gamma_+\gamma_-\over a_-}[-\cos(\varphi-\alpha)
-v_++v_-+v_+v_-\cos(\varphi-\alpha)],\nonumber \\
k^{\varphi}&=&{E\over a_-f_-}\sin(\varphi-\alpha),\quad
k^{\theta}=0\,.
\end{eqnarray}
This gives the boundary conditions for null geodesic equations in the inner 
region.
\end{enumerate}

In general, a local scattering angle is not well defined when the spatial 
curvatures of the outer and the inner parts are different. However, 
in the case of an empty void ($\rho^{\rm (m)}_-=0$), the inner region
 is described
 by a de Sitter space-time and the spatial curvature is negligible. 
Hence, for this case, we can introduce another Cartesian coordinates system
\begin{eqnarray}
ds^{2}=-dt_-^{2}+a_-^{2}(t_-)\left(dx_-^{2}+dy_-^{2}+dz_-^{2}\right)
\,,
\end{eqnarray}
and perform a coordinates transformation of $k^{\mu}$ from the spherical 
coordinates to the Cartesian coordinates as
\begin{eqnarray}
k^{x_-}&=&{E\over a_-}[-\cos\alpha+(v_--v_+)\cos\varphi]+O(v_{\pm}^2)
\nonumber \\
k^{y_-}&=&{E\over a_-}[-\sin\alpha+(v_--v_+)\sin\varphi]+O(v_{\pm}^2)
\,.
\end{eqnarray}
Finally we obtain the direction angle and the scattering angle as follows
\begin{eqnarray}
\tan\alpha_-&\equiv&{k^{y_-}\over k^{x_-}}
=\tan\alpha+(v_+-v_-){\sin(\varphi-\alpha)\over\cos^2\alpha}+O(v_{\pm}^2)
\nonumber \\
\Delta\alpha&\equiv&\alpha_--\alpha=(v_+-v_-)\sin(\varphi-\alpha)+O(v_{\pm}^2)
\,.
\end{eqnarray}

According to our numerical analysis, we find as follows:
\begin{itemize}
\item If $\Omega_{0}=1$, then $v_+\approx0.2HR$ and $v_-\approx-0.3HR$\,.
\item If $\Omega_{0}=0.3$, then $v_+\approx0.1HR$ and $v_-\approx-0.15HR$ 
\,.
\end{itemize}
Hence, we obtain the upper bound of the scattering angle, which is given in 
Eq.\,(\ref{Deltaalpha}).


\end{document}